# Interpretation of the coherency matrix for three-dimensional polarization states

José J. Gil

*Facultad de Educación. Universidad de Zaragoza. Pedro Cerbuna 12, 50009 Zaragoza, Spain.*
*ppgil@unizar.es*

**Abstract**

From an appropriate parameterization of the three-dimensional (3D) coherency matrix **R**, that characterizes the second-order, classical states of polarization, the coherency matrices are classified and interpreted in terms of incoherent decompositions. The relevant physical quantities derived from **R**, as the intensity, the degree of polarimetric purity, the indices of polarimetric purity, the angular momentum, the degree of directionality and the degree of linear polarization are identified and interpreted on the light of the case study performed. The information provided by **R** about the direction of propagation is clarified and it is found that coherency matrices with rank $\mathbf{R} = 2$, does not always represent states with a well-defined direction of propagation. Moreover, it is demonstrated the existence of 3D mixed states that cannot be decomposed into a superposition of a pure state, a 2D unpolarized state, and a 3D unpolarized state. Appropriate representation and interpretation for all the different types of 3D coherency matrices is provided through physical consistent criteria. Under the approach proposed, the conventional two-dimensional model arises naturally.

## 1. Introduction

A proper description of the polarization properties of electromagnetic waves relies on the concept of the coherency matrix, which is applicable regardless the particular band of the electromagnetic spectrum considered.

The study and characterization of three-dimensional (3D) states of polarization is a subject of high interest from both theoretical and experimental points of view. Several relevant contributions have been published in the recent years concerning aspects as the interpretation of physical quantities derived from the eigenvalues of the coherency matrix **R** [1-12]; geometric interpretation of 3D states [13-19]; coherent composition of pure (or totally polarized) states [20]; incoherent composition and decomposition of pure and mixed states [4,7,9,17,21]; 3D polarimetry [22,23]; statistical and coherence properties of 3D states [24-28], generalized Stokes parameters [16,1,9,13,29], etc.

However, additional effort is needed in order to get answers to questions such as: how to represent 3D states of polarization as combinations of states with a simple physical interpretation? Generally, unitary transformations of the coherency matrix do not correspond to rotations of the 3D laboratory reference frame, therefore, which unitary transformations are physically realizable? What kind of information about the propagation direction can be obtained from the coherency matrix? How to interpret 3D polarization states in terms of meaningful physical quantities? How many different physical situations can be distinguished and how to classify them?

The aim of this work is to provide appropriate and consistent responses to the previous questions by means of *1)* an adequate parameterization of the 3D coherency matrix **R**, as is the one proposed by Dennis [13] in terms of nine physical parameters; *2)* the case study of the different physical situations, identified through specific descriptors and analyzed by means of the arbitrary and characteristic decompositions of **R**, and *3)* the identification, definition and interpretation of parameters that provide complete and meaningful physical information.

**2. 3D states of polarization**

In the most general case, the three components of the electric field vector *E* of the electromagnetic wave should be considered in order to describe the evolution of *E*, which determines the polarization state. Let us consider a quasi-monochromatic wave of arbitrary form, propagating in an isotropic medium, and let $(\mathbf{e}_1, \mathbf{e}_2, \mathbf{e}_3)$ be a reference basis of orthonormal vectors along the respective axes *XYZ*. Given a point **r** in space, the analytic signals of the three components of *E* can be arranged as the *3D instantaneous Jones vector* [9]

$$\boldsymbol{\varepsilon}(t) = \begin{pmatrix} \varepsilon_1(t) \\ \varepsilon_2(t) \\ \varepsilon_3(t) \end{pmatrix} = \begin{pmatrix} A_x(t) \\ A_y(t) e^{i\delta_y(t)} \\ A_z(t) e^{i\delta_z(t)} \end{pmatrix} \quad (1)$$

In general, $\boldsymbol{\varepsilon}(t)$ has slow time dependence with respect to the coherence time, so that, for time intervals shorter than the coherence time, the polarization ellipse can be considered constant. For time intervals higher than the coherence time, the instantaneous Jones vector can vary, resulting in partially polarized states.

Note that totally-polarized (or *pure*) states are characterized by the fact that the quantities $A_y(t)/A_x(t)$, $A_z(t)/A_x(t)$, $\delta_y(t)$ and $\delta_z(t)$ are constant, and the 3D Jones vector (applicable only for pure states) is expressed without time dependence. It is well known that, in the case of pure states, the electric field describes an ellipse perpendicular to the direction of propagation (at point **r**). The 2D model is then easily reproduced by taking the direction of propagation as the transformed $Z_O$ reference axis, so that the third component of the 3D Jones vector vanishes.

The *coherency matrix* or *polarization matrix*, which contains all measurable second-order information about the state of polarization (including intensity) of an electromagnetic wave, is defined as the $3 \times 3$ Hermitian matrix $\mathbf{R} = \langle \boldsymbol{\varepsilon}(t) \otimes \boldsymbol{\varepsilon}^+(t) \rangle$ whose elements $r_{ij}$ are the second-order moments $r_{ij} = \langle \varepsilon_i(t) \varepsilon_j^*(t) \rangle$ $(i,j = 1,2,3)$ of the zero-mean analytic signals $\varepsilon_i(t)$, ($\varepsilon_i^*$ represents the complex conjugate of $\varepsilon_i$ and the brackets indicate time averaging over the measurement time). Thus, **R** is a covariance matrix and, therefore, its three eigenvalues $(\lambda_1, \lambda_2, \lambda_3)$ are non-negative. Moreover, **R** can be expressed as

$$\mathbf{R} = \begin{pmatrix} \langle A_x^2(t) \rangle & \langle A_x(t) A_y(t) e^{-i\delta_y(t)} \rangle & \langle A_x(t) A_z(t) e^{-i\delta_z(t)} \rangle \\ \langle A_x(t) A_y(t) e^{i\delta_y(t)} \rangle & \langle A_y^2(t) \rangle & \langle A_y(t) A_z(t) e^{i(\delta_y(t)-\delta_z(t))} \rangle \\ \langle A_x(t) A_z(t) e^{i\delta_z(t)} \rangle & \langle A_y(t) A_z(t) e^{-i(\delta_y(t)-\delta_z(t))} \rangle & \langle A_z^2(t) \rangle \end{pmatrix} \quad (2)$$

where the diagonal elements can be interpreted as the intensities associated with the respective *XYZ* components of the electric field and the total intensity (irradiance or power flux density of the wave, averaged over the measurement time) is given by

$$I = \operatorname{tr} \mathbf{R} = \langle A_x^2(t) \rangle + \langle A_y^2(t) \rangle + \langle A_z^2(t) \rangle. \tag{3}$$

As Fano pointed out [30], an appropriate basis for *nxn* coherency matrices is a set of Hermitian trace-orthogonal operators, that in the case of **R** is that constituted by the $3 \times 3$ identity matrix together with the eight Gell-Mann matrices (generators of the SU(3) group)

$$\boldsymbol{\omega}_{11} \equiv \begin{pmatrix} 1 & 0 & 0 \\ 0 & 1 & 0 \\ 0 & 0 & 1 \end{pmatrix}, \boldsymbol{\omega}_{12} \equiv \sqrt{\frac{3}{2}} \begin{pmatrix} 0 & 1 & 0 \\ 1 & 0 & 0 \\ 0 & 0 & 0 \end{pmatrix}, \boldsymbol{\omega}_{13} \equiv \sqrt{\frac{3}{2}} \begin{pmatrix} 0 & 0 & 1 \\ 0 & 0 & 0 \\ 1 & 0 & 0 \end{pmatrix},$$

$$\boldsymbol{\omega}_{21} \equiv \sqrt{\frac{3}{2}} \begin{pmatrix} 0 & -i & 0 \\ i & 0 & 0 \\ 0 & 0 & 0 \end{pmatrix}, \boldsymbol{\omega}_{22} \equiv \sqrt{\frac{3}{2}} \begin{pmatrix} 1 & 0 & 0 \\ 0 & -1 & 0 \\ 0 & 0 & 0 \end{pmatrix}, \boldsymbol{\omega}_{23} \equiv \sqrt{\frac{3}{2}} \begin{pmatrix} 0 & 0 & 0 \\ 0 & 0 & 1 \\ 0 & 1 & 0 \end{pmatrix}, \tag{4}$$

$$\boldsymbol{\omega}_{31} \equiv \sqrt{\frac{3}{2}} \begin{pmatrix} 0 & 0 & -i \\ 0 & 0 & 0 \\ i & 0 & 0 \end{pmatrix}, \boldsymbol{\omega}_{32} \equiv \sqrt{\frac{3}{2}} \begin{pmatrix} 0 & 0 & 0 \\ 0 & 0 & -i \\ 0 & i & 0 \end{pmatrix}, \boldsymbol{\omega}_{33} \equiv \frac{1}{\sqrt{2}} \begin{pmatrix} 1 & 0 & 0 \\ 0 & 1 & 0 \\ 0 & 0 & -2 \end{pmatrix}.$$

The Gell-Mann matrices have been normalized in order to ensure that their Euclidean norms coincide with that of the $3 \times 3$ identity matrix (in analogy to the 2D case, where the Euclidean norms of the Pauli matrices coincides with that of the $2 \times 2$ identity matrix). The notation used for these matrices is justified for the sake of simplicity as well as to emphasize the symmetry in some mathematical expressions [9]. Note that, leaving aside the normalization factor, the Pauli matrices are nested in the upper $2 \times 2$ submatrices of $\boldsymbol{\omega}_{12}, \boldsymbol{\omega}_{21}, \boldsymbol{\omega}_{22}$.

Thus, the coherency matrix **R** can be expanded as [3]

$$\mathbf{R} = \frac{1}{3} \sum_{i,j=1}^{3} q_{ij} \boldsymbol{\omega}_{ij}, \quad q_{ij} = \operatorname{tr}(\mathbf{R}\boldsymbol{\omega}_{ij}), \tag{5}$$

where the nine real coefficients $q_{ij}$ are the so-called *generalized Stokes parameters* [16].

A study of the 3D spectral density tensor devoted to plane waves was presented in Ref. [16], where it is shown that the real vector $\mathbf{v} \equiv (q_{32}, -q_{23}, q_{31})$ determines the plane of the polarization ellipse. Note that $q_{00} = \operatorname{tr} \mathbf{R} = I$, so that some expressions are simplified by using the normalized form $\hat{\mathbf{R}} \equiv \mathbf{R}/\operatorname{tr} \mathbf{R}$ of the coherency matrix.

Let us now consider the Euclidean norm and the trace norm of **R**, defined respectively as [9]

$$\|\mathbf{R}\|_2 \equiv \sqrt{\operatorname{tr}(\mathbf{R}^2)}; \quad \|\mathbf{R}\|_{tr} \equiv \operatorname{tr} \mathbf{R} = I, \tag{6}$$

The *3D degree of polarimetric purity* $P_{(3)}$ [9] can be defined as [31,32,1]

$$P_{(3)} = \sqrt{\frac{1}{2} \left( \frac{3 \|\mathbf{R}\|_2^2}{\|\mathbf{R}\|_{tr}^2} - 1 \right)}. \tag{7}$$

This invariant non-dimensional quantity is limited to the interval $0 \leq P_{(3)} \leq 1$. The upper limit corresponds to the case that **R** has only one nonzero eigenvalue (pure state). Conversely, $P_{(3)} = 0$ is reached when the three eigenvalues of **R** are equal (equiprobable mixture of states and zero correlation between the electric field components). For reasons explained in Ref. [9], we consider preferable to refer $P_{(3)}$ to as the degree of polarimetric purity rather than the degree of polarization.

Setälä et al. pointed out [1,2] that $P_{(3)}$ takes into account not only the purity of the mean polarization ellipse, but also the stability of the plane that contains the instantaneous components of the electric field of the wave. Nevertheless, $P_{(3)}$ does not provide complete information about the polarimetric purity of a 3D state of polarization **R**, and two quantities like the two *indices of polarimetric purity* [3,9,10,17] are required

$$P_1 = \frac{\lambda_1 - \lambda_2}{\text{tr}\mathbf{R}}, \ P_2 = \frac{\lambda_1 + \lambda_2 - 2\lambda_3}{\text{tr}\mathbf{R}}, \tag{8}$$

where $\lambda_1, \lambda_2, \lambda_3$ ($\lambda_1 \geq \lambda_2 \geq \lambda_3$) are the eigenvalues of **R**. These invariant non-dimensional parameters are restricted by the nested inequalities $0 \leq P_1 \leq P_2 \leq 1$.

Thus, while the pair $(P_1, P_2)$ provides detailed information of the structure of the polarimetric purity of **R**, $P_{(3)}$ represents an overall measure of polarimetric purity of **R**, which can be calculated from $P_1$ and $P_2$ by the weighted quadratic average [3,17]

$$P_{(3)} = \sqrt{3P_1^2 + P_2^2}/2. \tag{9}$$

By taking advantage of the analytic expressions for $(\lambda_1, \lambda_2, \lambda_3)$ obtained by Sheppard [18,19] from the *del Ferro-Cardano-Tartaglia-Vieta* solution for the cubic equation with real roots, the indices of polarimetric purity can be written as

$$P_1 = \frac{2}{\sqrt{3}} P_{(3)} \sin \psi, \ P_2 = 2 P_{(3)} \cos \psi;$$

$$\psi \equiv \frac{1}{3} \arccos\left(\frac{1 - 27 \det \hat{\mathbf{R}} - 3 P_{(3)}^2}{2 P_{(3)}^3}\right), \tag{10}$$

excluded the case $P_{(3)} = 0$, where $P_1 = P_1 = 0$. Note that the value of $\psi$ is limited by $0 \leq \psi \leq \pi/3$. The physically feasible region in the purity space $(P_1, P_2)$ has been studied and interpreted by us in previous papers [9,17], while interesting geometric representations for $(\lambda_1, \lambda_2, \lambda_3)$ and other derived quantities (including $P_1$ and $P_2$) have been presented by Sheppard [18,19].

The physical interpretation of $P_1, P_2$ and other parameters is considered in later sections on the light of the case study preformed.

## 3. Composition and decomposition of 3D states of polarization

### 3.1. Coherent composition

As a previous step to the study of the incoherent superposition of states of polarization, which must be performed through the additive composition of the respective coherency matrices, it is worth to consider the coherent composition of pure states, which must be realized through the additive composition of the respective Jones vectors.

Let us consider a point **r** in a linear, homogeneous and isotropic medium where two mutually coherent states, characterized by respective 3D Jones vectors $\boldsymbol{\varepsilon}_1$ and $\boldsymbol{\varepsilon}_2$ are superposed. The resultant state of polarization (at point **r**) is a pure state given by the Jones vector $\boldsymbol{\varepsilon} = \boldsymbol{\varepsilon}_1 + \boldsymbol{\varepsilon}_2$. The electric field of the combined state describes a well-defined ellipse, which, at point **r**, lies in the plane tangent to the wavefront [20].

The polarization ellipse of a pure state $\boldsymbol{\varepsilon}$ with nonzero ellipticity lies in a well-defined plane $\varPi$, and thus, the direction of propagation of the state (at point **r**) is well defined because it is necessarily perpendicular to $\varPi$. On the contrary, a 3D Jones vector representing a linearly polarized state is compatible with any direction of propagation perpendicular to the axis where the electric field of the electromagnetic wave lies. Consequently, a 3D Jones vector $\boldsymbol{\varepsilon}$ (and hence the corresponding coherency matrix) does not contain intrinsic information about the direction of propagation, but that information can be deduced from $\boldsymbol{\varepsilon}$, either as a fixed direction, for states with nonzero ellipticity, or as an arbitrary direction perpendicular to the polarization axis, for linearly polarized states (zero ellipticity). It will be worth keeping in mind this obvious result when we deal later with the incoherent composition and decomposition of linearly polarized states with coincident polarization axes.

Moreover, since the sum of Jones vectors is a Jones vector, given a 3D Jones vector $\boldsymbol{\varepsilon}$, it can always be expressed, in an infinite number of ways, as the sum of a number of 3D Jones vectors.

### 3.2 Arbitrary decomposition of the coherency matrix

Iterative [9] and constructive [21,32,34] general procedures for the decomposition of a coherency matrix into a convex sum of coherency matrices have been presented by us in previous works and are summarized below for the case of 3D polarization matrices.

Let us now consider the diagonalization $\mathbf{R} = \mathbf{U}\mathrm{diag}(\lambda_1, \lambda_2, \lambda_3)\mathbf{U}^\dagger$, where **U** is the unitary matrix whose columns are the eigenvectors of **R**, and $\mathrm{diag}(\lambda_1, \lambda_2, \lambda_3)$ represents the diagonal matrix composed of the ordered non-negative eigenvalues $(0 \leq \lambda_3 \leq \lambda_2 \leq \lambda_1)$.

As follows from Ref. [21] (where the case of four-dimensional covariance matrices representing material media is considered), **R** can be expressed as the following convex sum in terms of a set of $r$ (with $r \equiv \mathrm{rank}\,\mathbf{R}$) arbitrary independent 3D complex unit vectors $\mathbf{w}_i$ belonging to $\mathrm{Range}(\mathbf{R})$ (i.e., belonging to the subspace generated by the eigenvectors of **R** with nonzero eigenvalues)

$$\mathbf{R} = \sum_{i=1}^{r} p_i \mathbf{R}_i; \quad \mathbf{R}_i \equiv \mathrm{tr}\mathbf{R}\left(\mathbf{w}_i \otimes \mathbf{w}_i^\dagger\right); \quad p_i = \frac{1}{\mathrm{tr}\mathbf{H}\sum_{j=1}^{r}\frac{1}{\lambda_j}\left|\left(\mathbf{U}^\dagger \mathbf{w}_i\right)_j\right|^2}; \quad \sum_{i=1}^{r} p_i = 1, \tag{11}$$

where $|\ |$ indicates the modulus (or Euclidean norm).

Expansion (11)) provides the way for generating arbitrary complete sets of the coherency matrices $\mathrm{tr}\mathbf{R}\left(\mathbf{w}_i \otimes \mathbf{w}_i^\dagger\right)$ of the pure components as well as their corresponding coefficients $p_i$. Note that the minimum number of pure components of the arbitrary decomposition is equal to $r$. When $r = 3$, any arbitrary three-dimensional generalized complex basis $\mathbf{w}_i (i = 1, 2, 3)$ can be chosen. When $r = 2$, any arbitrary two-dimensional generalized complex basis $(\mathbf{w}_1, \mathbf{w}_2)$ belonging to $\mathrm{Range}(\mathbf{R})$ can be chosen. When $r = 1$, **R** represents a pure state (fixed polarization ellipse) and consequently the arbitrary decomposition has no physical interest, because it becomes a tautology. Hereafter, when appropriate to note that a state **R** is pure, we will denote its coherence matrix as $\mathbf{R}_p$.

By taking as $\mathbf{w}_i$ the eigenvectors $\mathbf{u}_i$ $(i=1,2,3)$ of $\mathbf{R}$, we get the following *spectral decomposition* of $\mathbf{R}$ as a particular case of the arbitrary decomposition

$$\mathbf{R} = \frac{\lambda_1}{\mathrm{tr}\mathbf{R}}\mathbf{U}\mathrm{diag}(\mathrm{tr}\mathbf{R},0,0)\mathbf{U}^+ + \frac{\lambda_2}{\mathrm{tr}\mathbf{R}}\mathbf{U}\mathrm{diag}(0,\mathrm{tr}\mathbf{R},0)\mathbf{U}^+ + \frac{\lambda_3}{\mathrm{tr}\mathbf{R}}\mathbf{U}\mathrm{diag}(0,0,\mathrm{tr}\mathbf{R})\mathbf{U}^+ \quad (12)$$

where each term in the sum is synthesized from the corresponding eigenvector $\mathbf{u}_i$ of $\mathbf{R}$ and its weight in the convex sum is proportional to the corresponding eigenvalue $\lambda_i$

$$\mathbf{R} = \sum_{i=1}^{3} \frac{\lambda_i}{\mathrm{tr}\mathbf{R}}\left[(\mathrm{tr}\mathbf{R})(\mathbf{u}_i \otimes \mathbf{u}_i^+)\right]. \quad (13)$$

It should be noted that, when one of the eigenvalues has a multiplicity higher than one, then the eigenvectors of the corresponding invariant subspace are not unique, and consequently the spectral decomposition is not unique.

By considering the possible values of the indices of purity, we observe that they have a direct link with the purity structure of $\mathbf{R}$ and provide more detailed information than the very value of $r \equiv \mathrm{rank}\,\mathbf{R}$. As a general overview, the following cases can be distinguished [17]: *1)* when $0 \leq P_1 \leq P_2 < 1$ $(r=3)$, $\mathbf{R}$ can be considered as composed of three pure states $\mathbf{R} = p_1\mathbf{R}_{p1} + p_2\mathbf{R}_{p2} + p_3\mathbf{R}_{p3}$; if, in particular, $P_2 = 0$ (and hence, $P_1 = 0$), $\mathbf{R}$ is proportional to the identity matrix, $\mathbf{R} = I\,\mathrm{diag}(1,1,1) \equiv \mathbf{R}_{u-3D}$, so that it represents a 3D unpolarized state (completely random polarization ellipse and completely random direction of propagation); *2)* when $0 \leq P_1 < P_2 = 1$ $(r=2)$, $\mathbf{R}$ can be considered as composed of two pure states $\mathbf{R} = p_1\mathbf{R}_{p1} + p_2\mathbf{R}_{p2}$; if, in particular $P_1 = 0$, $\mathbf{R}$ corresponds to a 2D unpolarized state propagating along a well-defined direction of propagation. Finally, when $P_1 = 1$ (and hence, $P_2 = 1$ and $r = 1$), $\mathbf{R}$ corresponds to a pure state.

### 3.3. Characteristic decomposition of the coherency matrix

While all the components of the arbitrary decomposition of $\mathbf{R}$ are pure states, it is also possible to decompose $\mathbf{R}$ by means of the following *characteristic* (or *trivial*) decomposition [9]

$$\mathbf{R} = \mathbf{U}\mathrm{diag}(\lambda_1,\lambda_2,\lambda_3)\mathbf{U}^+ = P_1 I \hat{\mathbf{R}}_1 + (P_2 - P_1) I \hat{\mathbf{R}}_2 + (1 - P_2) I \hat{\mathbf{R}}_3$$
$$\hat{\mathbf{R}}_1 \equiv \mathbf{U}\mathrm{diag}(1,0,0)\mathbf{U}^+,\ \hat{\mathbf{R}}_2 \equiv \frac{1}{2}\mathbf{U}\mathrm{diag}(1,1,0)\mathbf{U}^+,\ \hat{\mathbf{R}}_3 \equiv \frac{1}{3}\mathbf{U}\mathrm{diag}(1,1,1)\mathbf{U}^+ \quad (14)$$

where all the components have been chosen to have the same intensity $I = \mathrm{tr}\mathbf{R}$ and the respective coefficients of the components are expressed in terms of the indices of polarimetric purity $(P_1, P_2)$. Note that $\mathrm{rank}\,\hat{\mathbf{R}}_1 = 1$, $\mathrm{rank}\,\hat{\mathbf{R}}_2 = 2$ and $\mathrm{rank}\,\hat{\mathbf{R}}_3 = 3$.

Recall that the trivial decomposition of $\mathbf{R}$ cannot be always performed in the form of a sum of a pure state and a 3D unpolarized state. This is because, in general, pure $n \times n$ coherency matrices contain $2n-1$ independent parameters whereas *m*D-unpolarized $n \times n$ coherency matrices contain $2(n-m)+1$ independent parameters. In the case of a $2 \times 2$ coherency matrix, the well-known decomposition of a mixed state into a pure state and a 2D unpolarized state is retrieved.

The characteristic decomposition leads to a physical interpretation of $P_1$ as the ratio of power of the completely polarized part (or pure component) to the total power of the electromagnetic wave. Furthermore, $P_2$ is the relative portion of power once the completely random component has been subtracted.

## 4. The intrinsic coherency matrix

The physically realizable rotations of the laboratory reference frame *XYZ* are represented by $3\times 3$ orthogonal matrices $\mathbf{Q}$ (with $\det\mathbf{Q}=1$), so that the transformed coherency matrix $\mathbf{R}'$ representing the same state as $\mathbf{R}$ but referred to the new reference frame, is given by $\mathbf{R}'=\mathbf{Q}^T\mathbf{R}\mathbf{Q}$. Moreover, let us consider the decomposition of $\mathbf{R}$ into its real and imaginary parts $\mathbf{R}=\mathbf{R}_R+i\mathbf{R}_I$, where the real matrix $\mathbf{R}_R\equiv\mathrm{Re}(\mathbf{R})$ is symmetric and positive-semidefinite, whereas the imaginary matrix $\mathbf{R}_I\equiv\mathrm{Im}(\mathbf{R})$ is skew-symmetric $(\mathbf{R}_I=-\mathbf{R}_I^T)$. As Dennis pointed out [13], $\mathbf{R}_R$ can always be diagonalized through a particular rotation $\mathbf{Q}$ of the reference frame:

$$\mathbf{Q}^T\mathbf{R}_R\mathbf{Q}=\mathrm{diag}(a_1,a_2,a_3);\ (0\le a_3\le a_2\le a_1). \tag{15}$$

Thus, $\mathbf{R}_R$ defines an ellipsoid (called by Dennis the inertia ellipsoid), whose semiaxes $(a_1,a_2,a_3)$ are aligned along the respective transformed axes $X_O Y_O Z_O$ (Fig.1). Thus, $\mathrm{diag}(a_1,a_2,a_3)$ can be interpreted as the coherency matrix of a state composed of the incoherent superposition of three linearly polarized pure states

$$\begin{aligned}&\mathrm{diag}(a_1,a_2,a_3)=\mathbf{R}_{p1}+\mathbf{R}_{p2}+\mathbf{R}_{p3};\\ &\mathbf{R}_{p1}\equiv a_1\mathrm{diag}(1,0,0),\ \mathbf{R}_{p2}\equiv a_2\mathrm{diag}(0,1,0),\ \mathbf{R}_{p3}\equiv a_3\mathrm{diag}(0,0,1),\end{aligned} \tag{16}$$

with respective intensities $a_1$, $a_2$ and $a_3$. The previous decomposition is compatible with a variety of directions of propagation for each component (the only condition is that the direction of propagation is orthogonal to the respective polarization axis). For the sake of clarity in further physical interpretations, it results convenient the choice of the axis $Z_O$ as the common direction of propagation $\mathbf{k}$ for the pure states $\mathbf{R}_{p1}$ and $\mathbf{R}_{p2}$, while any axis orthogonal to $\mathbf{k}$ can be considered as the direction of propagation of the third pure component $\mathbf{R}_{p3}$ (Fig.1).

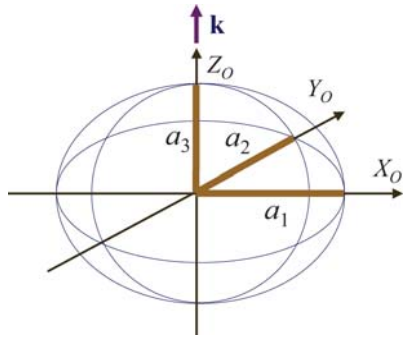

Fig. 1 (color online). Intensity ellipsoid, with semiaxes $a_1\ge a_2\ge a_3$, representing a mixed state constituted by the incoherent superposition of three pure linearly polarized states $\mathbf{R}_{p1}$, $\mathbf{R}_{p2}$ and $\mathbf{R}_{p3}$. The direction $\mathbf{k}$ along the reference axis $Z_O$ can be chosen as the common direction of propagation of the pure components $\mathbf{R}_{p1}$ and $\mathbf{R}_{p2}$ (with respective intensities $a_1, a_2$), whereas the third pure component $\mathbf{R}_{p3}$ (with intensity $a_3$), propagates along any direction orthogonal to **k**.

By applying the rotation $\mathbf{Q}$ to the whole matrix $\mathbf{R}$, we observe that the real and imaginary parts transform separately, so that we get the transformed coherency matrix [13]

$$\mathbf{R}_O \equiv \mathbf{Q}^T \mathbf{R} \mathbf{Q} = \mathbf{Q}^T \mathbf{R}_R \mathbf{Q} + i \mathbf{Q}^T \mathbf{R}_I \mathbf{Q} = \mathrm{diag}(a_1, a_2, a_3) + i\mathbf{R}'_I,$$

$$\mathbf{R}_O \equiv \begin{bmatrix} a_1 & -in_3 & in_2 \\ in_3 & a_2 & -in_1 \\ -in_2 & in_1 & a_3 \end{bmatrix}, \qquad (17)$$

where, as occurs for $\mathbf{R}_I$, $\mathbf{R}'_I$ is real and skew-symmetric. The orthogonal matrix $\mathbf{Q}$ can be expressed as a product of rotations around the respective axes *ZXY*

$$\mathbf{Q} = \mathbf{Q}_Z(\varphi)\mathbf{Q}_X(\alpha)\mathbf{Q}_Y(\beta) = \begin{bmatrix} \cos\varphi & -\sin\varphi & 0 \\ \sin\varphi & \cos\varphi & 0 \\ 0 & 0 & 1 \end{bmatrix} \begin{bmatrix} 1 & 0 & 0 \\ 0 & \cos\alpha & -\sin\alpha \\ 0 & \sin\alpha & \cos\alpha \end{bmatrix} \begin{bmatrix} \cos\beta & 0 & -\sin\beta \\ 0 & 1 & 0 \\ \sin\beta & 0 & \cos\beta \end{bmatrix}. \qquad (18)$$

Thus, $\mathbf{R}$ can be expressed as $\mathbf{R} = \mathbf{Q}\mathbf{R}_O\mathbf{Q}^T$, and it can be parameterized in the form proposed by Dennis [13] through the following nine independent parameters: the three orientation angles $(\varphi, \alpha, \beta)$; the semiaxes of the *intensity ellipsoid* (or *inertia ellipsoid* [13]) given by the *principal intensities* $(a_1, a_2, a_3)$, and the three components $(n_1, n_2, n_3)$, along the respective axes $X_O Y_O Z_O$, of the angular momentum **n** of the wave. In general, the orientation $\hat{\mathbf{n}} \equiv \mathbf{n}/\sqrt{n_1^2 + n_2^2 + n_3^2}$ of **n** differs from **k** [13]. The intensity *I* of the state $\mathbf{R}_O$ is given by

$$I = \mathrm{tr}\,\mathbf{R} = \lambda_1 + \lambda_2 + \lambda_3 = \mathrm{tr}\,\mathbf{R}_O = \mathrm{tr}\,\mathbf{R}_{p1} + \mathrm{tr}\,\mathbf{R}_{p2} + \mathrm{tr}\,\mathbf{R}_{p3} = a_1 + a_2 + a_3. \qquad (19)$$

Since the orthogonal similarity transformation (17) preserves the non-negativity (or convexity), the *intrinsic coherency matrix* $\mathbf{R}_O$ is positive semidefinite and, thus, the quantities $(a_1, a_2, a_3; n_1, n_2, n_3)$ must satisfy the following set of constraining inequalities [13] derived from the non-negativity of the leading principal minors of $\mathbf{R}_O$

$$a_1 \geq a_2 \geq a_3 \geq 0; \quad a_1 a_2 \geq n_3^2, a_1 a_3 \geq n_2^2, a_2 a_3 \geq n_1^2; \quad a_1 a_2 a_3 \geq a_1 n_1^2 + a_2 n_2^2 + a_3 n_3^2. \qquad (20)$$

The smaller is the third principal intensity $a_3$, the smaller is the solid angle around the axis $Z_O$ that limits the range of compatible orientations of **n**. When $a_3 = 0$, **n** is forced to lie along the axis $Z_O$, which, in turn, in this case is precisely the well-defined direction of propagation of the state $\mathbf{R}_O$.

Up to the limits set by the five restrictive inequalities (20), the quantities $(a_1, a_2, a_3; n_1, n_2, n_3)$ are independent and are intrinsic of a given coherency matrix **R**. It should be stressed that the fact that the only physically realizable unitary transformations of **R** are those that are orthogonal, leads to the indicated set of six physical parameters instead of the only three physical invariants derivable from the eigenvalues of **R**. In other words, not all the unitary transformations of **R** are physically realizable in the laboratory, and consequently a proper interpretation of the physical quantities involved in **R**, as well as an appropriate analysis of the arbitrary and characteristic decompositions of **R** must be performed through orthogonal transformations (and hence excluding the unitary transformations that are not orthogonal).

## 5. Case analysis

Once the arbitrary and characteristic decompositions, the structure of purity and the orthogonal transformations of a generic $3\times 3$ coherency matrix **R** have been considered, we are ready to undertake the study of the possible decompositions of a three-dimensional state of polarization represented by a given coherency matrix **R**. To perform a proper case analysis, it is necessary to pay attention to the values of the integer parameters $r \equiv \text{rank}\,\mathbf{R} = \text{rank}\,\mathbf{R}_O$ and $t \equiv \text{rank}\left[\text{Re}(\mathbf{R})\right] = \text{rank}\left[\text{Re}(\mathbf{R}_O)\right]$.

### 5.1. $\text{rank}\,\mathbf{R} = 1$.

In this case, the intrinsic coherency matrix $\mathbf{R}_O$ takes the form

$$\mathbf{R}_O = \begin{bmatrix} a_1 & -i\sqrt{a_1 a_2} & 0 \\ i\sqrt{a_1 a_2} & a_2 & 0 \\ 0 & 0 & 0 \end{bmatrix}, \quad (21)$$

so that, the electric field lies in the transformed plane $X_O Y_O$. Now, by introducing the parameters $s_0 \equiv a_1 + a_2$ and $s_3 \equiv 2\sqrt{a_1 a_2}$, together with the pair of parameters $s_1 \equiv (a_1 - a_2)\cos 2\theta$, $s_2 \equiv (a_1 - a_2)\sin 2\theta$, determined by the arbitrary choice of the angle $\theta$, we see that $\mathbf{R}_O$ can be expressed as [13]

$$\mathbf{R}_O = \frac{1}{2}\begin{bmatrix} s_0 + \sqrt{s_1^2 + s_2^2} & -is_3 & 0 \\ is_3 & s_0 - \sqrt{s_1^2 + s_2^2} & 0 \\ 0 & 0 & 0 \end{bmatrix}, \quad (22)$$

which corresponds to a 2D pure state propagating along the $Z_O$ axis. Furthermore, the polarization ellipse is oriented in such a manner that the major and minor semiaxes lie respectively in the new reference axes $X_O$ and $Y_O$ [note that this is a result of the effect of the matrix $\mathbf{Q}_Z(\varphi)$ in the orthogonal transformation (17)].

Thus, an additional rotation of an arbitrary angle $\theta$ around the axis $Z_O$ provides the general expression of the coherency matrix $\mathbf{R}_{pZ_O}$ of a pure state propagating along the direction $Z_O$ in terms of the Stokes parameters $(s_0, s_1, s_2, s_3)$ (Fig.2)

$$\mathbf{R}_{pZ_O} \equiv \mathbf{Q}_Z(\theta)\mathbf{R}_O \mathbf{Q}_Z^T(\theta) \equiv \frac{1}{2}\begin{bmatrix} s_0 + s_1 & s_2 - is_3 & 0 \\ s_2 + is_3 & s_0 - s_1 & 0 \\ 0 & 0 & 0 \end{bmatrix},$$

$$\left(\cos 2\theta \equiv \frac{s_1}{\sqrt{s_1^2 + s_2^2}}, s_0 = \sqrt{s_1^2 + s_2^2 + s_3^2}\right). \quad (23)$$

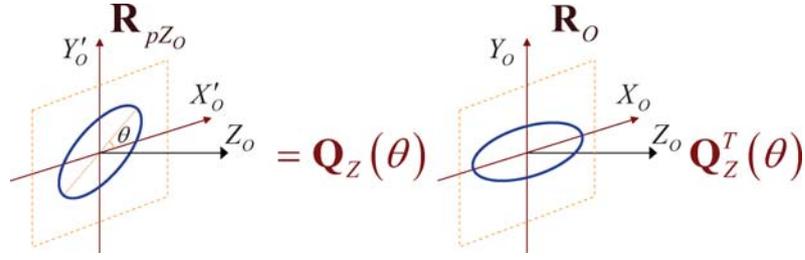

Fig. 2 (color online). Coherency matrix $\mathbf{R}_{pZ_O}$ of a generic pure state propagating along the axis $Z_O$, obtained from its *intrinsic coherency matrix* $\mathbf{R}_O$ through an arbitrary rotation $\theta$ around $Z_O$.

By considering now the two possible values of $t$ that are compatible with $r=1$, the following two sub-cases can be distinguished.

### 5.1.1. $r=1$, $t=1$, $(a_1 > 0,\ a_2 = a_3 = 0,\ n_1 = n_2 = n_3 = 0)$

$\mathbf{R}_O$ takes the simple form $\mathbf{R}_O = \mathrm{diag}(a_1, 0, 0)$, where $a_1$ can be interpreted as the Stokes parameter $s_1$ of a linearly polarized state whose polarization ellipse degenerates into a segment along the axis $X_O$. It should be noted that, in this case, $\mathbf{R}_O$ (and hence $\mathbf{R}$) is compatible with any direction of propagation perpendicular to the axis $X_O$. Obviously, under experimental conditions, it is common to have specific complementary information about the direction of propagation, but we stress that, in the case of a pure linearly polarized state, the very knowledge of the coherency matrix $\mathbf{R}$ does not determine the direction of propagation of the wave (Fig.3). Furthermore, we also note that, at the point in the space where $\mathbf{R}$ (and hence $\mathbf{R}_O$) is being considered, the incoherent superposition of a variety of pure states with linear polarizations along the $X_O$ axis, but with different directions of propagation, produces a pure state of linear polarization. Thus, without information additional to $\mathbf{R}$, this last case is polarimetrically indistinguishable from a pure state of linear polarization and fixed direction of propagation. The geometric nature of this peculiarity provides a method for superposing incoherently, at a fixed point in the space, a number of linearly polarized pure states whose arbitrary directions of propagation lie in a plane perpendicular to the common polarization axis $X_O$ and produce a pure state whose intensity is the sum of the intensities of the combined states.

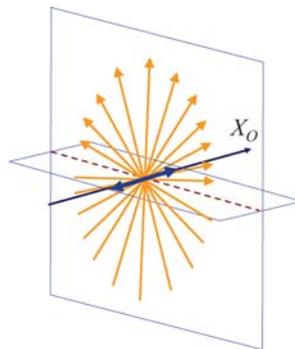

Fig. 3 (color online). Representation of a linearly polarized pure state as the incoherent superposition of an arbitrary number of linearly polarized states with the same polarization axis along the axis $X_O$, but with different directions of propagation

**5.1.2.** $r = 1$, $t = 2$, $\left(a_1 \geq a_2 > 0,\, a_3 = 0,\, a_1 a_2 = n_3^2,\, n_2 = n_3 = 0\right)$

In this case, the polarization ellipse determines a plane $X_O Y_O$, so that the propagation direction $Z_O$ is orthogonal to the said plane and therefore it is well defined. Once performed the orthogonal transformation of the reference axes, the case is reduced to a 2D pure state characterized by the coherency matrix

$$\mathbf{R}_{pZ_O} = \mathbf{Q}_Z(\theta) \mathbf{R}_O \mathbf{Q}_Z^T(\theta) = \frac{1}{2} \begin{bmatrix} s_0 + s_1 & s_2 - is_3 & 0 \\ s_2 + is_3 & s_0 - s_1 & 0 \\ 0 & 0 & 0 \end{bmatrix},$$

$$\left( \cos 2\theta \equiv \frac{s_1}{\sqrt{s_1^2 + s_2^2}},\; s_0 = \sqrt{s_1^2 + s_2^2 + s_3^2} \right).$$

(24)

**5.2.** rank $\mathbf{R} = 2$.

The interpretation of the coherency matrix $\mathbf{R}$ with rank $\mathbf{R} = 2$ depends substantially on the value of $t$. Since the value $t = 1$ is not compatible with $r = 2$, we distinguish the two possible cases $(r = 2, t = 2)$ and $(r = 2, t = 3)$.

**5.2.1.** $r = 2$, $t = 2$, $\left(a_1 \geq a_2 > 0,\, a_3 = 0,\, a_1 a_2 > n_3^2,\, n_2 = n_3 = 0\right)$

It is straightforward to show that, similarly as in the case $(r = 1, t = 2)$ the reference axes can be chosen in such a manner that the elements of the third row and of the third column of the coherency matrix are zero. The orthogonally transformed coherency matrix $\mathbf{R}_{Z_O}$ represents a partially polarized state with a well-defined direction of propagation $Z_O$, that is to say, a 2D partially polarized (or *mixed*) state

$$\mathbf{R}_{Z_O} = \mathbf{Q}_Z(\theta) \mathbf{R}_O \mathbf{Q}_Z^T(\theta) = \frac{1}{2} \begin{bmatrix} s_0 + s_1 & s_2 - is_3 & 0 \\ s_2 + is_3 & s_0 - s_1 & 0 \\ 0 & 0 & 0 \end{bmatrix},$$

$$\left( \cos 2\theta \equiv \frac{s_1}{\sqrt{s_1^2 + s_2^2}},\; s_0 > \sqrt{s_1^2 + s_2^2 + s_3^2} \right).$$

(25)

Once the direction of propagation $Z_O$ has been determined, the state of polarization can be considered two-dimensional and described by a generic $2 \times 2$ coherency matrix

$$\boldsymbol{\Phi}(I, P_1, \mathbf{u}) = I \frac{1}{2} \begin{bmatrix} 1 + P_1 u_1 & P_1(u_2 - iu_3) \\ P_1(u_2 + iu_3) & 1 - P_1 u_1 \end{bmatrix},$$

(26)

where the *intensity* $I = \operatorname{tr} \boldsymbol{\Phi}$ is the time averaged power density flux of the wave; $P_1$ (the first index of polarimetric purity) is the *degree of polarization*, and $u_i$ are the components of the unit vector $\mathbf{u}$ (Pauli axis) that summarizes the information about the azimuth $\varphi$ $(0 \leq \varphi < \pi)$ and ellipticity $\tan \chi$ $(-\pi/4 \leq \chi \leq \pi/4)$ of the average polarization ellipse

$$\mathbf{u} \equiv (u_1, u_2, u_3)^T = (\cos 2\chi \cos 2\varphi, \cos 2\chi \sin 2\varphi, \sin 2\chi)^T. \tag{27}$$

The state $\Phi(I, P_1, \mathbf{u})$, can also be represented through the corresponding *Stokes vector*

$$\mathbf{s} \equiv (s_0, s_1, s_2, s_3)^T = I \begin{bmatrix} 1 \\ P_1 \mathbf{u} \end{bmatrix}. \tag{28}$$

*5.2.1.1. Arbitrary decomposition.*

As it has been pointed out in previous works [7,9] a mixed 2D state $\Phi$ can always be considered as an incoherent composition of two totally polarized (or pure) states $\Phi_{p1}$ and $\Phi_{p2}$. One of them can arbitrarily be chosen and then the second one is totally determined by that choice, so that there exist infinite possibilities for decomposing $\Phi$ as a combination of two (or more) pure states

$$\Phi(I, P_1, \mathbf{u}) = p \Phi_{p1}(I, \mathbf{v}) + (1-p) \Phi_{p2}(I, \mathbf{w});$$
$$\Phi_{p1}(I, \mathbf{v}) \equiv I \frac{1}{2} \begin{bmatrix} 1+v_1 & v_2 - iv_3 \\ v_2 + iv_3 & 1-v_1 \end{bmatrix}, \quad \Phi_{p2}(I, \mathbf{w}) \equiv I \frac{1}{2} \begin{bmatrix} 1+w_1 & w_2 - iw_3 \\ w_2 + iw_3 & 1-w_1 \end{bmatrix}, \tag{29}$$
$$p = \frac{1-P_1^2}{2(1-P_1 \mathbf{u}^T \mathbf{v})}, \quad \mathbf{w} = \frac{P_1 \mathbf{u} - p\mathbf{v}}{1-p}.$$

A geometric view for the arbitrary decomposition of 3D states can be found in Ref. [7, p. 341]

Obviously, this arbitrary decomposition of $\Phi$ can be considered either: *a)* a convex combination of two pure states with equal intensities $I$, or *b)* an additive combination of two pure states $[p\Phi_{p1}]$ and $[(1-p)\Phi_{p2}]$ with respective intensities $(pI)$ and $(I-pI)$. Despite of the fact that both the said interpretations are equivalent and respectively physically realizable, it is particularly convenient to use the interpretation (a) because it is formulated in terms of representative states taken with the same trace norm $\|\Phi\|_{tr} = I$ [21].

As a particular case of the above arbitrary decomposition, the choice $\mathbf{v} = \mathbf{u}$ leads to the well-known spectral decomposition of $\Phi$ into two pure states represented by antipodal points on the Poincaré sphere

$$\Phi(I, P_1, \mathbf{u}) = \frac{1+P_1}{2} \Phi_p(I, \mathbf{u}) + \frac{1-P_1}{2} \Phi_{p2}(I, -\mathbf{u}), \tag{30}$$

Since the 2D states of polarization are usually represented by means of Stokes vectors, it is worth to bring up the corresponding expressions for the arbitrary and spectral decompositions

$$\mathbf{s} = I \begin{bmatrix} 1 \\ P_1 \mathbf{u} \end{bmatrix} = pI \begin{bmatrix} 1 \\ \mathbf{v} \end{bmatrix} + (1-p)I \begin{bmatrix} 1 \\ \mathbf{w} \end{bmatrix}, \quad \mathbf{s} = I \begin{bmatrix} 1 \\ P_1 \mathbf{u} \end{bmatrix} = \frac{1+P_1}{2} I \begin{bmatrix} 1 \\ \mathbf{u} \end{bmatrix} + \frac{1-P_1}{2} I \begin{bmatrix} 1 \\ -\mathbf{u} \end{bmatrix}. \tag{31}$$

We emphasize the potential applications of Eq. (29) in Stokes polarimetry (which has particular importance in several fields as, for instance, astronomic and atmospheric measurements [35]) because it provides a simple procedure for the polarimetric subtraction of a reference Stokes vector from that obtained by experimental measurements [21].

In summary, the state $\mathbf{R}_{Z_O}$ is expressed as an incoherent superposition of two pure states with the same direction of propagation $Z_O$ (Fig. 4)

$$\mathbf{R}_{Z_O} = pI\hat{\mathbf{R}}_{p1Z_O} + (1-p)I\hat{\mathbf{R}}_{p2Z_O}, \tag{32}$$

and can be interpreted in terms of the following six independent parameters:
- two orientation angles that determine the common direction of propagation of the two components;
- three stokes parameters of the pure state $I\hat{\mathbf{R}}_{pZ_O}(\mathbf{v})$,
- and the coefficient $p$ of $I\hat{\mathbf{R}}_{pZ_O}(\mathbf{v})$ in the convex sum.

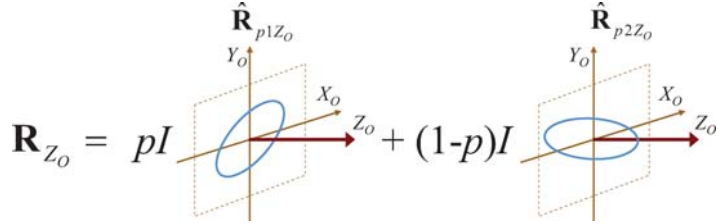

Fig. 4 (color online). Arbitrary representation of a 2D mixed state (i.e., $r=2, t=2$) as the incoherent superposition of two pure states with the same propagation direction $Z_O$.

### 5.2.1.2. Characteristic decomposition.

An alternative interpretation of a 2D mixed state is achieved through the corresponding characteristic decomposition

$$\boldsymbol{\Phi}(I, P_1, \mathbf{u}) = P_1 \boldsymbol{\Phi}_p(I, \mathbf{u}) + (1-P_1)\boldsymbol{\Phi}_u, \quad \boldsymbol{\Phi}_u \equiv \begin{bmatrix} 1 & 0 \\ 0 & 1 \end{bmatrix};$$
$$\mathbf{s} = I \begin{bmatrix} 1 \\ P_1 \mathbf{u} \end{bmatrix} = P_1 I \begin{bmatrix} 1 \\ \mathbf{u} \end{bmatrix} + (1-P_1)I \begin{bmatrix} 1 \\ \mathbf{0} \end{bmatrix}, \tag{33}$$

where $\boldsymbol{\Phi}_p$ represents a 2D pure state and $\boldsymbol{\Phi}_u$ represents a 2D unpolarized state, both propagating along the same direction $Z_O$. This is the well-known two-dimensional form of the characteristic decomposition.

Returning to the 3D representation, the characteristic decomposition is expressed as

$$\mathbf{R}_{Z_O} = P_1 \mathbf{R}_{pZ_O} + (1-P_1)\mathbf{R}_{u-2D};$$
$$\mathbf{R}_{pZ_O} \equiv I\frac{1}{2}\begin{bmatrix} 1+u_1 & u_2-iu_3 & 0 \\ u_2+iu_3 & 1-u_1 & 0 \\ 0 & 0 & 0 \end{bmatrix}, \quad \mathbf{R}_{u-2D} \equiv I\frac{1}{2}\begin{bmatrix} 1 & 0 & 0 \\ 0 & 1 & 0 \\ 0 & 0 & 0 \end{bmatrix}. \tag{34}$$

where $\mathbf{R}_{pZ_O}$ represents a pure state and $\mathbf{R}_{u-2D}$ represents a 2D unpolarized state, both propagating along the same direction $Z_O$; that is to say, once the laboratory reference axes have been appropriately rotated, the characteristic decomposition of a state $\mathbf{R}$ with $r=2, t=2$ is expressed as that of a generic 2D state [Eq. (33)].

Thus, in the case of $r=2, t=2$, the characteristic decomposition leads to the following interpretation of the coherency matrix in terms of six independent parameters (Fig.5):
- two orientation angles that determine the direction of propagation of the pure component $\mathbf{R}_{pZ_O}$;

- three stokes parameters of the pure component $\mathbf{R}_{pZ_O}$,
- and the degree of polarization $P_1$ of $\mathbf{R}_{Z_O}$, which is the coefficient of $\mathbf{R}_{pZ_O}$ in the convex sum.

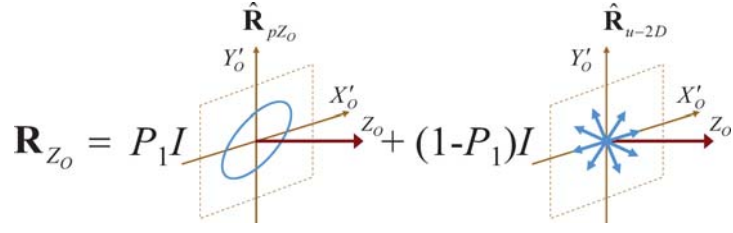

Fig. 5 (color online). Characteristic representation of a 2D mixed state (i.e., $r = 2$, $t = 2$) as the incoherent superposition of a pure state and a 2D unpolarized state with the same propagation direction $Z_O$.

**5.2.2.** $r = 2$, $t = 3$, $\left( a_1 \geq a_2 \geq a_3 > 0;\ a_1 a_2 a_3 = a_1 n_1^2 + a_2 n_2^2 + a_3 n_3^2 \right)$

When $t = 3$, the electric field $\boldsymbol{E}$ of the electromagnetic wave has necessarily three nonzero orthogonal components, so that $\boldsymbol{E}$ does not evolve inside a fixed plane, and thus the direction of propagation is not well defined. The analogy with the 2D representation is no longer applicable for the state represented by $\mathbf{R}$.

*5.2.2.1. Arbitrary decomposition.*

Any pure state belonging to $\text{Range}\,\mathbf{R}$ (note that $\text{Range}\,\mathbf{R}_O = \text{Range}\,\mathbf{R}$) can be considered as a component, and the second pure component as well as the respective coefficients, are easily determined. Thus, the arbitrary decomposition $\mathbf{R} = pI\,\hat{\mathbf{R}}_{p1} + (1-p)I\,\hat{\mathbf{R}}_{p2}$ (Fig. 6) can be performed either as indicated at the beginning of the present section, or through the following procedure [9]:

1. take an arbitrary 3D complex unit vector $\mathbf{w}_1$ belonging to $\text{Range}\,\mathbf{R}$, and synthesize the normalized coherency matrix $\hat{\mathbf{R}}_{p1} \equiv \mathbf{w}_1 \times \mathbf{w}_1^\dagger$ of the first component;
2. calculate the coefficient $p_1$ of $\hat{\mathbf{R}}_{p1}$: $p_1 = \left\{ \text{tr}\left[ \text{diag}(\lambda_1, \lambda_2, 0) \hat{\mathbf{R}}_{p1} \right] \right\}/I$; $(I \equiv \lambda_1 + \lambda_2)$,
3. and calculate the second pure component: $\hat{\mathbf{R}}_{p2} = \left( \hat{\mathbf{R}} - p\hat{\mathbf{R}}_{p1} \right)/(1-p)$.

Obviously, the previous procedure can be applied either to $\mathbf{R}$ or to $\mathbf{R}_O$ depending on the reference frame considered.

$$\mathbf{R} = pI \;\hat{\mathbf{R}}_{p1} + (1-p)I \;\hat{\mathbf{R}}_{p2}$$

Fig. 6 (color online). Arbitrary representation of a mixed state $\mathbf{R}$ with $\text{rank}\,\mathbf{R} = 2$ and $\text{rank}\left[\text{Re}(\mathbf{R})\right] = 3$, as the incoherent superposition of two pure states with different directions of propagation. In spite of $\text{rank}\,\mathbf{R} = 2$, $\mathbf{R}$ corresponds to a 3D mixed state.

Since any superposition of pure states with the same directions of propagation is represented by a coherency matrix $\mathbf{R}_O$ with $a_3 = 0$, we conclude that, when $r = 2, t = 3$, the state $\mathbf{R}$ can be considered as the superposition of two pure states whose propagation directions are different. In other words, the previous analysis demonstrates the surprising result that a coherency matrix $\mathbf{R}$ with $\text{rank}\,\mathbf{R} = 2$ can be synthesized through the superposition of two pure states propagating along different directions (provided $\text{rank}\,\mathbf{R}_O = 3$).

Consequently, in the case of $r = 2, t = 3$, the coherency matrix can be interpreted in terms of the following set of eight independent parameters:

- two orientation angles that determine the direction of propagation of the first component $\hat{\mathbf{R}}_{p1}$ of the arbitrary decomposition;
- three stokes parameters of the pure state $\hat{\mathbf{R}}_{p1}$;
- two orientation angles that determine the direction of propagation of the second pure component $\hat{\mathbf{R}}_{p2}$ of the arbitrary decomposition,
- and the coefficient $p_1$ of the convex sum in the arbitrary decomposition of $\mathbf{R}$.

*5.2.2.2. Characteristic decomposition.*

In this case, the characteristic decomposition of $\mathbf{R}$ is formulated as follows

$$\mathbf{R} = \mathbf{U}\begin{bmatrix} \lambda_1 & 0 & 0 \\ 0 & \lambda_2 & 0 \\ 0 & 0 & 0 \end{bmatrix}\mathbf{U}^\dagger = P_1 \mathbf{R}_p + (1-P_1)\mathbf{R}_m;$$

$$\mathbf{R}_p \equiv I\,\mathbf{U}\begin{bmatrix} 1 & 0 & 0 \\ 0 & 0 & 0 \\ 0 & 0 & 0 \end{bmatrix}\mathbf{U}^\dagger, \quad \mathbf{R}_m \equiv \frac{I}{2}\mathbf{U}\begin{bmatrix} 1 & 0 & 0 \\ 0 & 1 & 0 \\ 0 & 0 & 0 \end{bmatrix}\mathbf{U}^\dagger,$$

(35)

in terms of the pure component $\mathbf{R}_p$ and the non-pure component $\mathbf{R}_m$. While $\mathbf{R}_p$ has an immediate physical interpretation as a completely polarized state, the physical interpretation of $\mathbf{R}_m$ requires the consideration of the value of the integer parameter $r_m \equiv \text{rank}\left[\text{Re}(\mathbf{R}_m)\right]$:

*a)* when $r_m = 2$, $\mathbf{R}_m$ represents a 2D unpolarized state $\mathbf{R}_{u-2D}$ propagating along a direction different from that of $\mathbf{R}_p$, and therefore $\mathbf{R}$ can be interpreted by means of the following eight parameters provided by the characteristic decomposition of $\mathbf{R}$ (Fig. 7):

- two orientation angles that determine the direction of propagation of the pure component $\mathbf{R}_p$;
- three stokes parameters of the pure component $\mathbf{R}_p$;
- two orientation angles that determine the direction of propagation of the 2D unpolarized component $\mathbf{R}_m \equiv \mathbf{R}_{u-2D}$,
- and the first index of polarimetric purity (or degree of polarization) $P_1$ of $\mathbf{R}$, which is the coefficient of $\mathbf{R}_p$ in the convex sum (we recall that, when $r < 3$, the second index of polarimetric purity $P_2$ is equal to one, and thus the coefficient $P_2 - P_1$ of $\mathbf{R}_m$ becomes $1 - P_1$).

*b)* when $r_m = 3$, $\mathbf{R}_m$ represents a 3D partially polarized (but not totally random) state without a well-defined direction of propagation. The state $\mathbf{R}_m$ corresponds to the case (2b) studied previously. From the characteristic decomposition of $\mathbf{R}$, it can be interpreted by means of the following eight parameters (Fig. 8):

- two orientation angles that determine the direction of propagation of the pure component $\mathbf{R}_p$;
- three stokes parameters of the pure component $\mathbf{R}_p$,
- and the three principal intensities of the electric field of $\mathbf{R}_m$, i.e., the eigenvalues of $\mathrm{Re}(\mathbf{R}_m)$. Note that, with this choice, the value of $P_1$ is obtainable from the indicated set of parameters.

Fig. 7 (color online). Characteristic representation of a mixed state $\mathbf{R}$ with $\mathrm{rank}\,\mathbf{R} = 2$, $\mathrm{rank}\left[\mathrm{Re}(\mathbf{R})\right] = 3$ and $\mathrm{rank}\left[\mathrm{Re}(\mathbf{R}_m)\right] = 2$, as the incoherent superposition of a pure state and a 2D unpolarized state, with different directions of propagation. In spite of $\mathrm{rank}\,\mathbf{R} = 2$, $\mathbf{R}$ corresponds to a 3D mixed state.

Fig. 8 (color online). Characteristic representation of a mixed state $\mathbf{R}$ with $\mathrm{rank}\,\mathbf{R} = 2$, $\mathrm{rank}\left[\mathrm{Re}(\mathbf{R})\right] = 3$ and $\mathrm{rank}\left[\mathrm{Re}(\mathbf{R}_m)\right] = 3$, as the incoherent superposition of a pure state and a 3D mixed state. In spite of $\mathrm{rank}\,\mathbf{R} = 2$, $\mathbf{R}$ corresponds to a 3D mixed state.

**5.3.** $\operatorname{rank}\mathbf{R}=3$, $\left(a_1 \geq a_2 \geq a_3 > 0,\ a_1 a_2 a_3 > a_1 n_1^2 + a_2 n_2^2 + a_3 n_3^2\right)$.

In this case, the only achievable value of $t \equiv \operatorname{rank}\left[\operatorname{Re}(\mathbf{R}_O)\right]$ is $t=3$, so that the electric field $E$ of the electromagnetic wave has necessarily three nonzero orthogonal components, and thus the direction of propagation is not well defined. As in the previous case, in order to get appropriate physical interpretation, let us consider separately the arbitrary and characteristic decompositions.

*5.3.1. Arbitrary decomposition*

When $\operatorname{rank}\mathbf{R}=3$, the arbitrary decomposition of $\mathbf{R}$ can be performed either as indicated at the beginning of the present section, or through the following procedure [9]:

1. take an arbitrary 3D complex unit vector $\mathbf{w}_1$ (note that $\mathbf{w}_1$ necessarily belongs to $\operatorname{Range}\mathbf{R}$, because $\operatorname{Range}\mathbf{R}$ covers the complete 3D complex space) and synthesize the normalized coherency matrix $\hat{\mathbf{R}}_{p1} \equiv \mathbf{w}_1 \times \mathbf{w}_1^\dagger$ of the first component;

2. calculate the coefficient $p_1$ of $\hat{\mathbf{R}}_{p1}$:
$$p_1 = \left\{\operatorname{tr}\left[\operatorname{diag}(\lambda_1,\lambda_2,\lambda_3)\hat{\mathbf{R}}_{p1}\right]\right\}/I;\ (I \equiv \lambda_1 + \lambda_2 + \lambda_3);$$

3. calculate the remainder coherency matrix $\hat{\mathbf{R}}_r = \left[\hat{\mathbf{R}} - p_1\left(\mathbf{w}_1 \times \mathbf{w}_1^\dagger\right)\right]/(1-p_1)$;

4. take an arbitrary 3D complex unit vector $\mathbf{w}_2$ belonging to $\operatorname{Range}\hat{\mathbf{R}}_r$ and synthesize the normalized coherency matrix $\hat{\mathbf{R}}_{p2} \equiv \mathbf{w}_2 \times \mathbf{w}_2^\dagger$ of the second component;

5. calculate the coefficient $p_2$ of $\hat{\mathbf{R}}_{p2}$ in the convex sum
$$p_2 = \frac{1}{I}\operatorname{tr}\left\{\operatorname{diag}(\lambda_1,\lambda_2,\lambda_3)\left[\mathbf{w}_2 \otimes \mathbf{w}_2^\dagger\right]\right\},$$

6. and calculate the third pure component and its coefficient through the expressions:
$$\hat{\mathbf{R}}_{p3} = \left[\hat{\mathbf{R}} - p_1\left(\mathbf{w}_1 \times \mathbf{w}_1^\dagger\right) - p_2\left(\mathbf{w}_2 \times \mathbf{w}_2^\dagger\right)\right]/(1 - p_1 - p_2),\quad p_3 = 1 - p_1 - p_2.$$

With respect to the directions of propagation of the three pure components of the arbitrary decomposition, there are two possibilities: *1)* two components have the same direction of propagation, but different to that of the remainder component, and *2)* the three pure components have different directions of propagation.

The arbitrary decomposition of $\mathbf{R}$ leads to its interpretation in terms of the following nine independent parameters (Fig. 9):

- two orientation angles that determine the direction of propagation of the first component $\hat{\mathbf{R}}_{p1}$;

- three stokes parameters of the pure state $\hat{\mathbf{R}}_{p1}$;

- two orientation angles that determine the direction of propagation of the pure component $\hat{\mathbf{R}}_{p2}$ (or, if the said direction coincides with the direction of propagation of $\hat{\mathbf{R}}_{p1}$, the orientation angles that determine the direction of propagation of $\hat{\mathbf{R}}_{p3}$),

- and the coefficients $p_1$ and $p_2$ (recall that $p_3 = 1 - p_1 - p_2$).

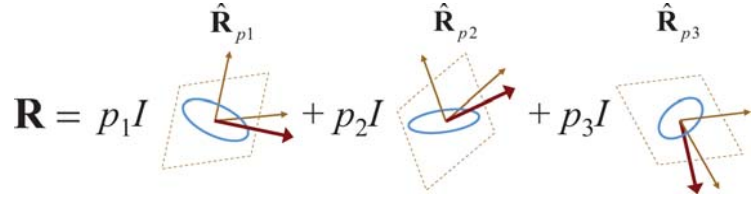

Fig. 9 (color online). Arbitrary representation of a mixed state **R** with $\operatorname{rank}\mathbf{R} = 3$ as the incoherent superposition of three pure states with different directions of propagation.

### 5.3.2. Characteristic decomposition

When $\operatorname{rank}\mathbf{R} = 3$, the characteristic decomposition of **R** has the general form

$$\mathbf{R} = \mathbf{U}\begin{bmatrix} \lambda_1 & 0 & 0 \\ 0 & \lambda_2 & 0 \\ 0 & 0 & \lambda_3 \end{bmatrix}\mathbf{U}^\dagger = P_1\mathbf{R}_p + (P_2 - P_1)\mathbf{R}_m + (1 - P_2)\mathbf{R}_{u-3D};$$

$$\mathbf{R}_p \equiv I\mathbf{U}\begin{bmatrix} 1 & 0 & 0 \\ 0 & 0 & 0 \\ 0 & 0 & 0 \end{bmatrix}\mathbf{U}^\dagger, \quad \mathbf{R}_m \equiv \frac{I}{2}\mathbf{U}\begin{bmatrix} 1 & 0 & 0 \\ 0 & 1 & 0 \\ 0 & 0 & 0 \end{bmatrix}\mathbf{U}^\dagger, \quad \mathbf{R}_{u-3D} \equiv \frac{I}{3}\begin{bmatrix} 1 & 0 & 0 \\ 0 & 1 & 0 \\ 0 & 0 & 1 \end{bmatrix}.$$

(36)

$\mathbf{R}_p$ represents a pure state and $\mathbf{R}_{u-3D}$ represents a 3D unpolarized state ($P_2 = 0$). The characteristic decomposition is completed with the second component $\mathbf{R}_m$, whose interpretation can be performed as follows in terms of the value of the auxiliary parameter $r_m \equiv \operatorname{rank}[\operatorname{Re}(\mathbf{R}_m)]$.

*a)* when $r_m = 2$, $\mathbf{R}_m$ represents a 2D unpolarized state $\mathbf{R}_{u-2D}$ propagating along a direction different than that of $\mathbf{R}_p$. Thus, **R** can be expressed as

$$\mathbf{R} = P_1\mathbf{R}_p + (P_2 - P_1)\mathbf{R}_{u-2D} + (1 - P_2)\mathbf{R}_{u-3D},$$ (37)

and therefore, **R** can be interpreted by means of the following eight independent parameters (Fig. 10):

- two orientation angles that determine the direction of propagation of the pure component $\mathbf{R}_p$;
- three stokes parameters of the pure component $\mathbf{R}_p$;
- two orientation angles that determine the direction of propagation of the 2D unpolarized component $\mathbf{R}_m \equiv \mathbf{R}_{u-2D}$,
- and the two indices of polarimetric purity $(P_1, P_2)$ of **R**.

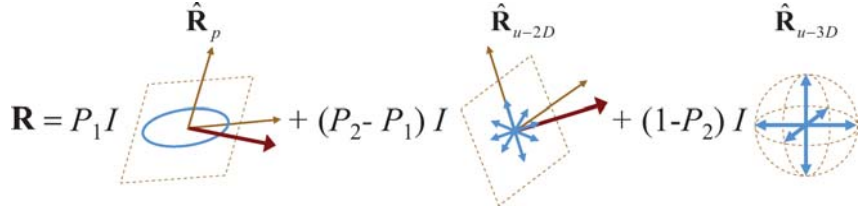

Fig. 10 (color online). Characteristic representation of a mixed state **R** with rank $\mathbf{R} = 3$, and rank$\left[\text{Re}(\mathbf{R}_m)\right] = 2$, as the incoherent superposition of a pure state, a 2D unpolarized state (with different direction of propagation) and a 3D unpolarized state.

b) when $r_m = 3$, the components $\mathbf{R}_p$ and $\mathbf{R}_m$ can be recombined and re-decomposed in such a manner that the characteristic decomposition of **R** can be rewritten in the following appropriate form

$$\mathbf{R} = \frac{P_1 + P_2}{2} I\hat{\mathbf{R}}_{p1} + \frac{P_2 - P_1}{2} I\hat{\mathbf{R}}_{p2} + (1 - P_2) I\hat{\mathbf{R}}_{u-3D};$$

$$\hat{\mathbf{R}}_{p1} \equiv \mathbf{U}\begin{bmatrix} 1 & 0 & 0 \\ 0 & 0 & 0 \\ 0 & 0 & 0 \end{bmatrix}\mathbf{U}^\dagger, \hat{\mathbf{R}}_{p2} \equiv \frac{1}{2}\mathbf{U}\begin{bmatrix} 0 & 0 & 0 \\ 0 & 1 & 0 \\ 0 & 0 & 0 \end{bmatrix}\mathbf{U}^\dagger, \hat{\mathbf{R}}_{u-3D} \equiv \frac{1}{3}\begin{bmatrix} 1 & 0 & 0 \\ 0 & 1 & 0 \\ 0 & 0 & 1 \end{bmatrix}, \quad (38)$$

where the two first components are pure states with different directions of propagation and $\hat{\mathbf{R}}_{u-3D}$ is a 3D unpolarized state. Thus, **R** can be interpreted by means of the following nine independent parameters (Fig. 11):

- two orientation angles that determine the direction of propagation of the first pure component $\mathbf{R}_{p1}$;
- three stokes parameters of $\mathbf{R}_{p1}$;
- two orientation angles that determine the direction of propagation of the second pure component $\mathbf{R}_{p2}$,
- and the two indices of purity $(P_1, P_2)$ of **R**.

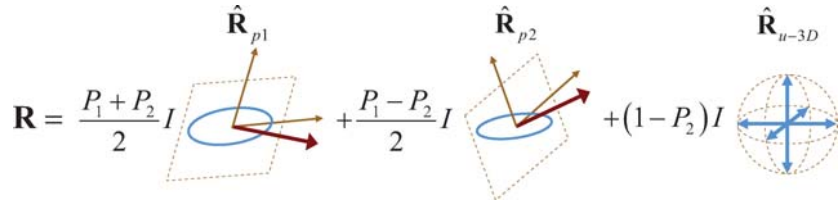

Fig. 11 (color online). Characteristic representation of a mixed state **R** with rank $\mathbf{R} = 3$, and rank$\left[\text{Re}(\mathbf{R}_m)\right] = 3$ as the incoherent superposition of two pure states (with different directions of propagation) and a 3D unpolarized state.

## 6. The degree of directionality

As it has been pointed out above, the 3D coherency matrix **R** is defined for a given fixed point **r** in the space, and does not contain direct information about the direction of propagation of the electromagnetic wave. Nevertheless, in the previous section we have found that, when $t \equiv rank\left[\operatorname{Re}(\mathbf{R}_O)\right] = 2$, the evolution of the electric field defines a fixed plane $\Pi$, and consequently, the direction **k** perpendicular to that plane, and hence perpendicular to the wavefront, can properly be considered as the direction of propagation at point **r**. That is to say, the 3D state of polarization given by **R** is polarimetrically indistinguishable from a plane wave with the same coherency matrix **R** and propagating along the direction **k**.

Moreover, when $t = 3$, the evolution of the electric field of the wave defines the intensity ellipsoid, so that, except for the case of 3D unpolarized states, the direction **k** perpendicular to the plane of symmetry containing the maximum ellipse (Fig. 1), can be considered as a the average direction of propagation. Thus, it is highly desirable to define a measure of the directional purity of the state **R**, i.e., a measure of the distance from the directionality of **R** to the random directionality of a 3D unpolarized state. As a previous attempt to get a definition of such a *degree of directionality*, we proposed the use of the second index of polarimetric purity $P_2$, but the analysis performed in the previous sections shows that, when $r \equiv \operatorname{rank} \mathbf{R} = 2$ and $t = 3$, then $P_2 = 1$, despite the fact that the direction of propagation is not well defined. On the light of the interpretation of the intrinsic coherency matrix $\mathbf{R}_O(\mathbf{R})$ and inspired by the definition of $P_2$ in terms of the eigenvalues of **R**, we find that the desired definition of the degree of directionality $P_d$ is given by the nondimensional quantity

$$P_d \equiv \frac{a_1 + a_2 - 2a_3}{a_1 + a_2 + a_3}, \tag{39}$$

where, as indicated in Eq. (15), the real, non-negative, quantities $(a_1, a_2, a_3)$, are defined from the diagonalization $\mathbf{Q}^T \mathbf{R}_R \mathbf{Q} = \operatorname{diag}(a_1, a_2, a_3)$ of the real part $\mathbf{R}_R$ of **R**, with the choice $0 \le a_3 \le a_2 \le a_1$. The degree of directionality $P_d$ reaches its maximum value $P_d = 1$ for $a_3 = 0$. Moreover $P_d = 0$ corresponds exclusively to a 3D unpolarized state $\mathbf{R}_{u-3D}$, whose associated direction of propagation is completely random (i.e., the intensity ellipsoid is a sphere). In agreement with the case study performed above, intermediate cases have appropriate and consistent values in the interval $0 < P_d < 1$. Thus, in our opinion, $P_d$ is a proper measure of the degree of directionality.

The particular case of a pure linearly polarized state ($a_2 = a_3 = 0$) is out of the range of application of $P_d$. In fact, in this limiting case $P_d = 1$, while the corresponding coherency matrix can be obtained trough the incoherent superposition of an arbitrary set of linearly polarized pure states propagating along different directions, provided that such directions lie in the plane perpendicular to the polarization axis.

By taking advantage of the analogies between the expressions derived from the eigenvalues $(\lambda_1, \lambda_2, \lambda_3)$ of **R** and that derived from the eigenvalues $(a_1, a_2, a_3)$ of $\operatorname{Re}(\mathbf{R})$, we define the degree of linear polarization $P_l$ as the nondimensional parameter

$$P_l \equiv \frac{a_1 - a_2}{a_1 + a_2 + a_3}, \tag{40}$$

whose possible values are limited by $0 \le P_l \le P_d$ ($P_d \le 1$). The minimum value $P_l = 0$ ($a_1 = a_2$) corresponds to a state with an intensity ellipsoid of revolution with semiaxes $(a_1, a_1, a_3)$, so that its intersection with the plane perpendicular to the average direction of propagation **k** is a circumference of radio $a_1$. Regardless the value of $P_d$, $P_l = 0$ corresponds to states with equal principal intensities in the plane perpendicular to **k**. When, in particular, $P_d = 0$, the intensity ellipsoid is a sphere and corresponds to a 3D unpolarized state. The maximum value $P_l = 1$, entails $P_d = 1$ and corresponds to a pure state of linear polarization. In the case of 2D states $(P_d = 1)$, $P_l = \sqrt{s_1^2 + s_2^2}/s_0$, as corresponds to the natural and commonly used definition of the degree of linear polarization. Thus, the inspection of the values of $P_l$ for all the possible 3D states of polarization shows that $P_l$ gives an appropriate and consistent measure of the degree of linear polarization.

Regardless the intrinsic physical meaning of $(a_1, a_2, a_3)$ as the semiaxes of the intensity ellipsoid, equivalent physical information can be represented through the set of quantities $(I, P_l, P_d)$, whose physical meaning is particularly appropriate for the study and analysis of three-dimensional states of polarization.

## 7. Conclusions

At a given point in the space, the second-order state of polarization of an arbitrary electromagnetic wave is characterized by means of the corresponding coherency matrix **R**, which can be interpreted in terms of a set of nine well-defined parameters, namely [13], the principal intensities $(a_1, a_2, a_3)$; the angular momentum $\mathbf{n} \equiv (n_1, n_2, n_3)$, and the set of three orientation angles $(\varphi, \alpha, \beta)$. This parameterization constitutes an adequate framework for the analysis and interpretation of any three-dimensional state of polarization through the corresponding arbitrary and characteristic decompositions.

Together with the integer parameter $r \equiv \text{rank}\, \mathbf{R}$, which determines the minimum number of pure components in the arbitrary decomposition, we have found that the integer parameter $t \equiv \text{rank}\left[\text{Re}(\mathbf{R})\right]$ plays a fundamental role in the physical interpretation of the possible three-dimensional states of polarization. $t = 1$ corresponds to pure states with linear polarization. In this case, as indicated in the previous section, the direction of propagation is not determined by **R**, even though it is true that the experimentalists have usually complementary information enough to determine it. When $t = 2$, the state **R** is reduced to a conventional two-dimensional state of polarization, with a well-defined direction of propagation; thus, the arbitrary and the characteristic decompositions of **R** translate into that of two-dimensional states and, in particular, the arbitrary decomposition is necessarily composed of two pure states with the same directions of propagation; moreover, the characteristic decomposition has the well-known form of a superposition of a pure state and a unpolarized two-dimensional state. In the case that $t = 3$, even if $r = 2$, it is no longer possible to assign a well-defined direction of propagation to the state **R**, so that at least two of the three pure arbitrary components have different propagation directions; moreover, the characteristic decomposition adopts particular forms depending on the value of an additional auxiliary integer parameter. The case study performed clarifies the interpretation and the role played by the set of two indices of polarimetric purity $(P_1, P_2)$ as physically invariant quantities that give nondimensional appropriate measures of the structure of purity of a state **R**, beyond the overall information provided by the degree of polarimetric purity $P_{(3)} = \sqrt{3P_1^2 + P_2^2}/2$.

Inspired by the expression of the second index of polarimetric purity $P_2$ in terms of the eigenvalues of **R**, the degree of directionality $P_d$ has been defined, which gives an appropriate and consistent measure of the stability of the direction of propagation of the state of polarization represented by **R**.

The transformation of the coherency matrix performed through the appropriate rotation of the reference frame, provides the intrinsic coherency matrix $\mathbf{R}_O$ characterized by the set of six parameters $(a_1, a_2, a_3; n_1, n_2, n_3)$. The physical information contained in $\mathbf{R}_O$ can also be represented through the following alternative set of meaningful parameters: intensity $I$, degree of polarization $P_1$, second index of polarimetric purity $P_2$ (i.e., the relative portion of the wave obtained once the 3D unpolarized portion has been subtracted); the degree of directionality $P_d$, the degree of linear polarization $P_l$ and the magnitude $n$ of the angular momentum **n**.

In summary, the approach presented, we think, constitutes a useful tool for the study, representation and interpretation of the complete variety of three-dimensional states of polarization in terms of appropriate and well-defined physical parameters.

**Acknowledgement**

This research was supported by the Spanish Ministry of Economía y Competitividad, Grant FIS2011-22496, and by the Gobierno de Aragón, group E99.